\documentclass[11pt]{article}

\usepackage{xcolor}
\usepackage{enumitem} 
\usepackage{graphicx}
\usepackage{epsfig}
\usepackage{amsfonts}
\usepackage{amssymb}
\usepackage{amsmath,amssymb}

\usepackage{url}
\usepackage{subcaption}  
 \usepackage[utf8]{inputenc}
\DeclareMathOperator\arctanh{arctanh}

\newcommand{\ee}{\end{equation}}
\newcommand{\eea}{\end{eqnarray}}
\newcommand{\be}{\begin{equation}}
\newcommand{\bea}{\begin{eqnarray}}

\def\Bitm{\begin{itemize}}
\def\Eitm{\end{itemize}}
\def\Blist#1#2{\begin{list}{#1}{\parsep=0pt \itemsep=0pt%   
  \listparindent=0pt #2}}
\def\Elist{\end{list}}
\long\def\ignore#1#2{\def\ignoreflag{#1}\long\def\tmptext{#2}
  \ifnum\ignoreflag>1 #2 \fi}

\usepackage{comment}
\usepackage
[
a4paper,% other options: a3paper, a5paper, etc
left=1.8cm,
right=1.8cm,
top=3cm,
bottom=4cm,
]{geometry}

\begin{document}

\title{\bf Spinning extremal dyonic black holes  
\\
 in  $\gamma=1$
Einstein-Maxwell-dilaton theory }
 \vspace{1.5truecm}
 \date{\today}
%\pacs{04.20.Jb, 04.40.Nr}

\author{
	{\large Jose Luis Bl\'azquez-Salcedo}$^{1}$,  
	{\large Carlos Herdeiro}$^{2}$,  
	{\large Eugen Radu}$^{2}$,
	\\
	{\large Etevaldo dos Santos Costa Filho}$^{2}$ 
	and
	{\large Kunihito Uzawa}$^{3,4}$
	\vspace*{0.2cm}
	\\
	$^{1}${\small Departamento de F\'isica Te\'orica and IPARCOS, Universidad Complutense de Madrid, E-28040 Madrid, Spain}
	\\
	$^{2}${\small Departamento de Matem\'atica da Universidade de Aveiro and } 
	\\ {\small Centre for Research and Development in Mathematics and Applications (CIDMA),} 
	\\ {\small Campus de Santiago, 3810-183 Aveiro, Portugal}  
	\\ $^{3}$ {\small Department of Physics, School of Science and Technology, Kwansei Gakuin University,} 
	\\ {\small Sanda, Hyogo 669-1337, Japan} 
	\\ $^{4}$ {\small Research and Education Center for Natural Sciences, Keio University,} 
	\\ {\small Hiyoshi 4-1-1, Yokohama, Kanagawa 223-8521, Japan}
}

\maketitle

\begin{abstract}  
We propose a general framework for the study of asymptotically flat 
spinning dyonic {\it extremal} black holes (eBHs)
in $D=4$ Einstein-Maxwell-dilaton  theory. 
Restricting to the stringy value $\gamma=1$
of the dilaton coupling constant, we report on the existence of a one parameter family of eBHs which are free 
of pathologies, provided their
magnetic and electric charges are equal.
An understanding of this condition is found from
a study of the near horizon limit of the 
solutions, both perturbative closed form  
and numerical solutions being presented.

\end{abstract}

%\tableofcontents

%%%%%%%%%%%%%%%%%%%%%%%%%%%%%%%%%%%%%%%%%%%%%%%%%%%%%%%%%%%%%%%%%%%%%%%%%%%%%%
\section{Introduction  }
%%%%%%%%%%%%%%%%%%%%%%%%%%%%%%%%%%%%%%%%%%%%%%%%%%%%%%%%%%%%%%%%%%%%%%%%%%%%%%

The physics of the extremal Black Holes (eBHs) has 
 received much interest in the last decades, associated with various developments in gravity and high energy physics.
For example, the first successful statistical
counting of BH entropy
was performed for an extremal static
Reissner-Nordstr\"om (RN) BH in $D=5$ spacetime dimensions \cite{Strominger:1996sh}, then generalized to its spinning counterpart~\cite{Breckenridge:1996is,Herdeiro:2000ap}. 

The majority of studies in the literature restrict themselves to gravity models in which the eBHs
are known in a closed form.
The archetypal example here is the $D=4$ electrovacuum model, with the Kerr-Newman extremal dyonic solution (KN) \cite{Newman:1965my} providing the most general
(physical) configuration.
This eBH possesses four global charges:
the mass $M$, angular momentum $J$, electric charge $Q$,
and magnetic charge $P$, subject to a specific relation
as imposed by the zero Hawking temperature condition.
 As originally noticed in \cite{Bardeen:1999px}, 
 the Kerr and KN eBH solutions admit a decoupled near-horizon limit,
which contains a $AdS_2$ factor and has  
 an enhanced $SO(2, 1) \times U(1)$ isometry.
 This observation has led to the emergence of the Kerr/CFT correspondence~\cite{Guica:2008mu},
 a duality which provides a
microscopic framework for understanding the BH entropy (see Ref. \cite{Compere:2012jk} for a review).

Although some extensions of the extremal KN BH have been reported
in the literature
(in particular in a number of supersymmetric theories),
relatively little is known about
its generalization to other models. 
%well motivated 
In particular, the (nonperturbative) generalization  of the  KN BH 
in Einstein-Maxwell-dilaton (EMd) theory is  known in closed form\footnote{ 
A $\gamma=1$ generalization  of the KN BH in the low energy
effective field theory for heterotic string theory has been reported in 
the Ref. \cite{Sen:1992ua},  although for a more general model than   (\ref{action}).
EMd exact solutions with arbitrary $\gamma$
are known  in the slow-rotation
\cite{Horne:1992zy}
or weakly charged approximations \cite{Casadio:1996sj}. {We also note that static, purely electrically charged solutions were studied in \cite{GIBBONS1988741,Garfinkle:1990qj}, whereas static dyonic solutions were investigated in \cite{Kallosh:1992ii,Cremonini:2023vwf}. 
   }
} 
only for the Kaluza-Klein (KK) value of the 
 dilaton coupling constant $\gamma$ 
\cite{Rasheed:1995zv,Larsen:1999pp},
numerical solutions\footnote{Non-extremal 
rotating dyonic BHs with 
$\gamma \neq (0,\sqrt{3})$ were studied numerically in~\cite{Kleihaus:2003df}.} 
with $P=0$
being reported recently~\cite{Herdeiro:2025blx}
for several values of $\gamma$.
At the same time,
some
recent results indicate that extremal BHs with
regular horizons are rather an exception
\cite{Horowitz:2024kcx}.
As discussed in \cite{Herdeiro:2025blx}, the 
(ellectrically) charged spinning EMd BHs with $0<\gamma < \sqrt{3}$ 
possess a zero temperature limit, which, albeit regular in terms of curvature invariants, exhibits a $pp$-singularity.

It would be interesting to see if the picture found in Ref.  
\cite{Herdeiro:2025blx}
changes in the presence of a net magnetic charge, $P\neq 0$.
Indeed, this is the case for static EMd solutions: while the extremal limit 
of the purely electric solutions is singular,
the dyonic BH solutions possess a smooth extremal limit.
There, the case 
\begin{eqnarray}
\label{condition}
    P^2=Q^2,
\end{eqnarray}
is special, with a vanishing dilaton field, the solution
coinciding with the  Reissner-Nordstr\"om eBH.

This suggests that {\it the spinning solutions satisfying the
condition (\ref{condition}) may also possess special properties},
a conjecture which,
at least for $\gamma=1$,
is confirmed by the results in this work.
While the generic dyonic configurations possess all numerical issues
found in the purely electric case \cite{Herdeiro:2025blx},
we provide evidence for the existence of a one parameter family 
of regular eBHs, provided that the condition (\ref{condition})
is satisfied.

Some understanding of this result
is achieved by studying the (decoupled) near horizon limit 
of the solutions.
There, the metric and the matter functions at the horizon have an angular dependence,
being found by solving a set of ordinary differential
equations, the condition (\ref{condition}) emerging when studying perturbative solutions.

The problem is tackled by using both analytical and numerical methods. 
In particular, we propose a general
framework for the numerical study
of both spinning dyonic BHs in EMd theory together 
and the associated near horizon configurations 
(although the displayed numerical results are for the stringy case $\gamma=1$).

This paper is organized as follows. In Section~\ref{sec2} the EMd model is presented and the (numerical) extremal solutions are described. Section~\ref{sec3} discusses the near horizon solutions and Section~\ref{sec4} provides our conclusions and final remarks.

%%%%%%%%%%%%%%%%%%%%%%%%%%%%%%%%%%%%%
\section{The extremal charged, spinning 
black holes  
}
\label{sec2}
%%%%%%%%%%%%%%%%%%%%%%%%%%%%%%%%%%%%%% 

%%%%%%%%%%%%%%%%%%%%%%%%%%%%%%%%%%%%%%%%%%%%%
\subsection{The framework}
%%%%%%%%%%%%%%%%%%%%%%%%%%%%%%%%%%%%%%%%%%%%%

We consider the $D=4$
EMd  action
%(we set $c=G= 1$) 
\begin{eqnarray}
\label{action}
\mathcal{S}= \frac{1}{16 \pi}\int d^4 x \sqrt{-g} 
\left(R-2\partial_\mu \phi\partial^\mu \phi 
- e^{-2\gamma \phi } F_{\mu\nu}F^{\mu\nu} \right),
\end{eqnarray}
where $R$ is the Ricci scalar, $F_{\mu \nu}=\partial_\mu A_\nu-\partial_\nu A_\mu$ is the Maxwell field 
and $\phi$ is the scalar field;
also, $\gamma$ is an input parameter which characterizes the coupling of the dilaton
with the Maxwell field.
The corresponding equations of motion are
\begin{eqnarray}
\label{eqEinstein}
&&
E_{\mu\nu} \equiv
R_{\mu\nu} - \frac{1}{2}Rg_{\mu\nu} 
 - 2\left[\partial_{\mu} \phi\partial_{\nu}\phi 
-\frac{1}{2}g_{\mu\nu}\partial_{\rho}\phi\partial^{\rho}\phi 
+ e^{-2\gamma \phi }\left(F_{\mu\rho}F_{\nu}^{~\rho} 
- \frac{1}{4}g_{\mu\nu} F_{\rho\sigma}
F^{\rho\sigma} 
\right)
  \right] 
  =0\ 
  ,~~~
 \\
 &&
\label{eqScalar}
\frac{1}{\sqrt{-g}}\partial_{\mu}(\sqrt{-g}\partial^{\mu}\phi) 
=  -\frac{\gamma}{2}\ e^{-2\gamma \phi }
F_{\rho\sigma}F^{\rho\sigma} \   ,
~~
%&&
%\label{eqM}
\partial_{\mu}(\sqrt{-g} e^{-2\gamma \phi }F^{\mu\nu})= 0 \ .
\end{eqnarray}

We consider stationary, axially symmetric BH
spacetimes with two Killing vector fields, 
%$\xi=\partial_t$ 
$\partial_t$
and 
$\partial_\varphi$
%$\eta=\partial_\varphi$.
The considered general framework
proposed 
in~\cite{Herdeiro:2025blx}
can straightforwardly  be extended to include a magnetic charge, together with an extremal horizon.
As discussed there, the BHs in EMd theory obeys the circularity condition,
the metric ansatz containing three undetermined
functions.
A suitable parametrization
which accommodates the presence of an
extremal horizon reads
%The metric Ansatz is constructed to accommodate the presence of an
%extremal horizon, with a line-element
%
\begin{equation}
\label{line_BH}
d s^2=-e^{2 F_0(r,\theta)} N(r)^2d t^2+e^{2 F_1(r,\theta)}\left(\frac{d r^2}{N(r)^2}+r^2 d \theta^2\right)+e^{-2 F_0(r,\theta)} r^2 \sin ^2 \theta\left(d \varphi-W(r,\theta) d t\right)^2\,,
\end{equation}
where
\begin{equation}
N(r) = 1-\dfrac{r_H}{r}\,,
\end{equation}
the event horizon being located at a surface with constant radial
variable $r = r_H >0$;
also  $(r, \theta,\varphi)$ are spheroidal
coordinates with the usual range.
The gauge and scalar fields are parametrized by
\begin{equation}
\label{matter}
 {A}_\mu d x^\mu=\left[ A_t(r,\theta)
 - {A}_{\varphi}(r,\theta) W(r,\theta) \right]d t+ {A}_{\varphi}(r,\theta) d \varphi\,,\qquad\qquad \phi\equiv \phi(r,\theta)\, .
\end{equation}
As such,
the field  eqs. (\ref{eqEinstein}), (\ref{eqScalar})
results in six partial
differential equations,
which possess no general-$\gamma$ closed form solution, being solved numerically
subject to proper boundary conditions\footnote{For $\gamma=(0,\sqrt{3})$,
the known EMd exact solutions can be written in the form (\ref{line_BH}), (\ref{matter}).}.

The boundary conditions at the event horizon and also the numerical treatment of the problem are
simplified by introducing a new radial coordinate
\begin{equation}
x=\sqrt{r^2-r_H^2},
\end{equation}
such that the line element (\ref{line_BH})
becomes
\begin{eqnarray}
&&
\nonumber
ds^2=-\frac{x^4 e^{2 F_0(x,\theta)}   }{u(x) H(x)^2}dt^2 
+e^{2 F_1(x,\theta)}\left(\frac{H(x)^2}{x^2}dx^2+u(x) d \theta^2\right)
+e^{-2 F_0(x,\theta)} u(x) \sin ^2 \theta\left(d \varphi-W(x,\theta) d t\right)^2\,,
\\
&&
\label{line_BHn}
\qquad \qquad {\rm with}~
u(x)=x^2+r_H^2,~~H(x)=r_H+\sqrt{x^2+r_H^2}~.
\end{eqnarray}

For the solutions to approach  at spatial infinity 
a Minkowski spacetime background, we require
$ F_0= F_1= W=0$,  
 while the conditions for the 
 matter fields are $\phi=0$,
 together with
%~nonzero magnetic flux as implied by %the condition
$A_\varphi=P\cos \theta$
and
${\rm lim}_{x\to \infty}x^2\partial_x A_t=-Q$. In our approach 
the electric and magnetic
charges are input parameters. 

On the symmetry axis, 
we impose for $\theta=0$ the boundary conditions
$\partial_\theta F_0=\partial_\theta F_1=\partial_\theta W=\partial_\theta A_t=\partial_\theta\phi=0$,
$A_\varphi=P $,
which coincide with those for
$\theta=\pi$, except for 
$A_\varphi=-P$. In addition, the absence of
a conical singularity imposes $F_1+F_0=0$
at $\theta=0,\pi$.

The boundary conditions at the horizon, $x=0$, are more subtle. 
 Assuming the existence there of a Taylor expansion of the solution, 
 %on finds the following expression
 a direct computation yields
\begin{eqnarray}
\label{expansion-eh}
&&
F_1= f_{10}(\theta)+x^2 f_{12}(\theta)+\dots,~~
F_0= f_{00}(\theta)+x^2 f_{02}(\theta)+\dots,~~
W= \Omega_H+x^2 w_{(2)}+\dots,~~
\\
&&
\phi= \phi_{0}(\theta)+x^2 \phi_{2}(\theta)+\dots,~~
A_\varphi= a_{\varphi{(0)}}(\theta)+x^2 a_{\varphi{(2)}}(\theta)+\dots,~~
A_t= x^2 a_{t(2)}+\dots,~~
\end{eqnarray}
where  $\Omega_H$ is
the event horizon angular velocity
and $a_{t(2)}$ and $w_{(2)}$ are undetermined constants.
Also, one can show that 
the constraint equation $E_r^\theta$ implies
%\begin{eqnarray}
$
 f_{00}(\theta)=f_{10}(\theta)+\alpha,
 $
%\end{eqnarray}
with $\alpha$ a constant.
The functions 
$f_{10}(\theta)$,
$\phi_{0}(\theta)$
 $a_{\varphi{(0)}}(\theta)$
 and 
 $f_{12}(\theta)$,
 $f_{02}(\theta)$,
 $\phi_{2}(\theta)$
 and
 $a_{\varphi{(2)}}(\theta)$
 are solutions of a complicated set of 
 ordinary differential equations 
 (ODEs)
 (see the discussion in Section 3.2).
 As such, at $x=0$
 we impose
 $\partial_x F_0= \partial_x F_1=\partial_x \phi=
 \partial_x A_\varphi=A_t=0$
 %\partial_x A_t=0$
 and
 $W=\Omega_H$.

The ADM mass $M$ and the angular momentum $J$
of the solutions
can be read from the asymptotic sub-leading behaviour
of the metric functions:
 	\begin{equation}\label{ADM_metric}
		g_{t t}=-1+\frac{2 M}{r}+\ldots, \quad g_{\varphi t}=-\frac{2 J }{r} \sin ^2 \theta+\ldots,
	\end{equation}
while the electric charge $Q$,  magnetic charges $P$
and the dilaton charge $D$
enter  the asymptotics of
    the matter functions:
 \begin{equation}
A_t = V_0+\frac{Q }{r}+\ldots, \quad A_{\varphi} =P \cos\theta +\ldots\,~     \phi= \dfrac{D}{r}+\cdots\,,
\end{equation}
with $V_0$
the electrostatic potential.

As usual,
the total angular momentum $J$ can be expressed 
as the sum of 
the event horizon and bulk contributions \cite{Wald:1984rg},
 \begin{eqnarray}
\label{J1}
J=J_{H}+
\frac{1}{8\pi}
%\int e^{-2 \gamma \phi}
\int R_\varphi^t \sqrt{-g} d^3x~,
\end{eqnarray}
an expression which can be simplified
by using the field equations and
 noticing that  $T_\varphi^t $
can be written as a total divergence,
$T_\varphi^t\sqrt{-g} =(\sqrt{-g} A_\varphi F^{\mu t})_{,\mu}$.
Then, for the  employed ansatz
and asymptotics, 
the  equation
 eq. (\ref{J1})
results in 
\begin{eqnarray}
\label{J}
&&
 J= 
\int_0^\pi d\theta
\frac{e^{-4 F_0}u^{3/2}\sin \theta}{8x }
\bigg(
-u \sin^2 \theta W_{,x}
+ 4 e^{2 F_0 -2 \gamma \phi}
A_\varphi (A_{t,x}-A_\varphi W_{,x})
\bigg)
\Big|_{x=0}~,
\end{eqnarray}
the contribution of the boundary term at infinity vanishing for the employed boundary conditions.
From the Maxwell equations, the electric charge
can also be written as a horizon integral,
 \begin{eqnarray}
\label{Q}
Q=
\int_0^\pi d\theta
\frac{u^{3/2}\sin \theta}{x}
e^{-2(F_0+\gamma \phi)}
(
 A_{t,x}-A_\varphi W_{,x}
 )
\Big|_{x=0}~.
\end{eqnarray}

 As for extremal KN BHs, the EMd eBHs have a topologically spherical horizon.  
 Geometrically,   the horizon is a deformed sphere,
with an  area
\begin{eqnarray}
\label{AH}
&&
A_H=2\pi r_H^2 \int_0^\pi d\theta \sin \theta~e^{F_1(0,\theta)-F_0(0,\theta)}  \ .
\end{eqnarray}

The extremal EMd BHs satisfy the mass formula
\begin{equation}
M=%\frac{\kappa A}{4 \pi}+
2 \Omega_H J+2 V_0 Q_e+\frac{D}{\gamma}~.
\end{equation}

%%%%%%%%%%%%%%%%%%%%%%%%%%%%%%%%%%%%%%%%%%%%%
\subsection{The solutions}
%%%%%%%%%%%%%%%%%%%%%%%%%%%%%%%%%%%%%%%%%%%%%

In  the numerics, we have found useful to
introduce a compactified radial
coordinate 
$X = x/(1 + x)$,
which maps the semi-infinite region $[0, \infty)$ to the finite region $[0, 1]$,  avoiding the use of a cutoff radius\footnote{Therefore the boundary condition at infinity $(X=1)$ for the electric potential reads 
$\partial_X A_t=-Q$, while $V_0$ is read from the numerical output; $r_H>0$ is an arbitrary input  parameter, with $r_H=0.1$
for most of the reported solutions.}.
The framework in the previous Section is valid for any value of the 
dilaton coupling constant $\gamma$.
However, in what follows 
we shall present results
for 
the stringy value $\gamma=1$, only\footnote{Similar results have been found for $\gamma=0.5$
and $\gamma=1.5$.}. 

The  numerical calculations are performed by
using  a professional software package \cite{SCHONAUER1989279,SCHONAUER1990279,SCHONAUER2001473}
that employs a finite difference method with an arbitrary grid and arbitrary consistency order. 
%T
For most of the computations have been
 utilized an equidistant grid with 251 points in $X$, covering the integration region $0 \leq X \leq 1$, and 61 points in $\theta$ direction 
 (with $0 \leq \theta \leq \pi$).
The Newton-Raphson method is employed in numerics, which requires a good first guess to start a successful iteration procedure.

 Our starting configuration  typically 
consisted in the extremal Kerr BH (eK),
which is a solution of the field equations
(with $Q=P=0$)
for any $\gamma$.
In principle,
the extremal dyonic configurations can be found by increasing in small steps
the parameters $P,Q$ which are fixed via boundary conditions.
However, the resulting generic configurations are plagued by some issues,
 similar to those noticed in \cite{Herdeiro:2025blx} for the zero temperature limit of
the EMd solutions with $P=0$.
For example, the overall numerical accuracy was rather low,
despite the fact that the functions which enter the Ansatz
(\ref{line_BH}),
(\ref{matter}) 
have smooth profiles,
and no indication is found for the existence of a divergent
behaviour of the
Kretschmann  and Ricci scalars.
An observed issue is that all configurations with
$P^2\neq Q^2$
possess a parallely propagated ($pp$) curvature
singularity: tidal forces as felt
by a timelike observer infalling into the BH diverge as the extremal horizon is approached \cite{Horowitz:1997ed}.
Furthermore, we have noticed that the radial second derivative of the
scalar field is not smooth enough, appearing to possess an oscillatory behaviour close to the horizon.
We shall not report further on such $generic$ configurations.

The situation changes
for the {\it extremal}  configurations obeying 
(\ref{condition})
in which case we have found a continuous set of solutions
which are free of the above mentioned issues.
As in the static case, these solutions 
also possess a vanishing dilaton charge, $D=0$, but now the scalar field is nonzero in the presence of rotation.

\begin{figure}[h!]
	\makebox[\linewidth][c]{%
		\begin{subfigure}[b]{8cm}
			\centering 
\includegraphics[width=8cm]{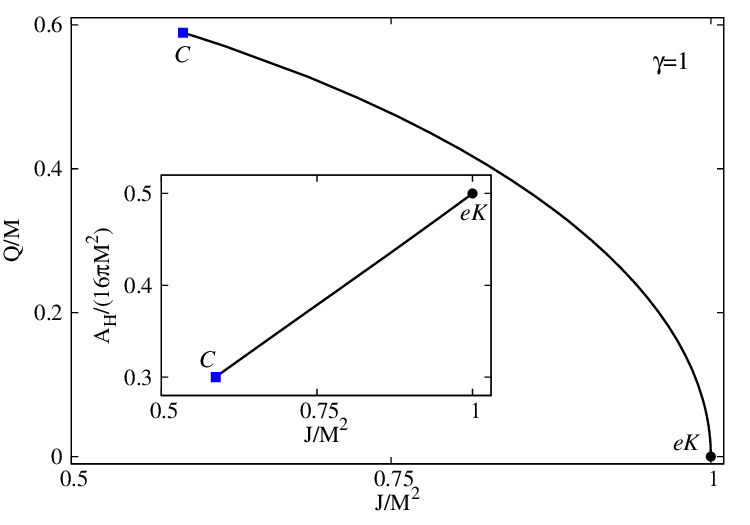}
		\end{subfigure}%
		\begin{subfigure}[b]{7.85cm}
			\centering 
\includegraphics[width=7.85cm]{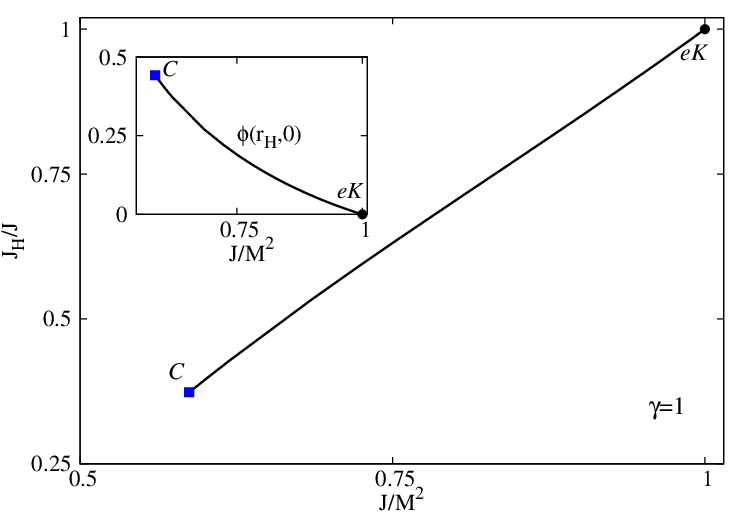}
		\end{subfigure}%
        }
\caption{\small  
{\it Left:}
The electric charge  and the event horizon area
(inset)
are shown as a function of angular momentum 
for $\gamma=1$ extremal BH solutions
(all quantities are given in units of mass).
{\it Right:}
The ratio between horizon angular momentum and total angular momentum
together with the value of the scalar field at $\theta=0$ on the horizon
are shown as a function of angular momentum.
}
\label{BH-data}
\end{figure}

The resulting family of regular solutions
smoothly emerges from Kerr eBH and
can be parametrized 
by the reduced
angular momentum $j=J/M^2$. 
Along it, for  given mass, 
the angular 
momentum decreases, as well as the 
event horizon area, 
while the electric (magnetic) charge increase,
as displayed in the Figure \ref{BH-data}.
%%%%%%%%%%%%%%%%%%%%%%%%%%%%%%%%%%%%%%%%%%%%%%%%%%%%%%%%%%%%%%%%%%%%%%%%%%%%%%%%%%%%%%%%%%%%%%%%%%%%%%%%%%%%%%%%%%%%%%%%%%%%%%%
%%%%%%%%%%%%%%%%%%%%%%%%%%%%%%%%%%%%%%%%%%%%%%%%%%%
% {\small \hspace*{3.cm}{\it  } }
%%%%%%%%%%%%%%%%%%%%%%%%%%%%%%%%%%%%%%%%%%%%%%%%%%%%%%%%%%%%%%%%%%%%%%%%%%%%%%
%

\begin{figure}[h]
	\makebox[\linewidth][c]{%
		\begin{subfigure}[b]{8cm}
			\centering 
\includegraphics[width=8cm]{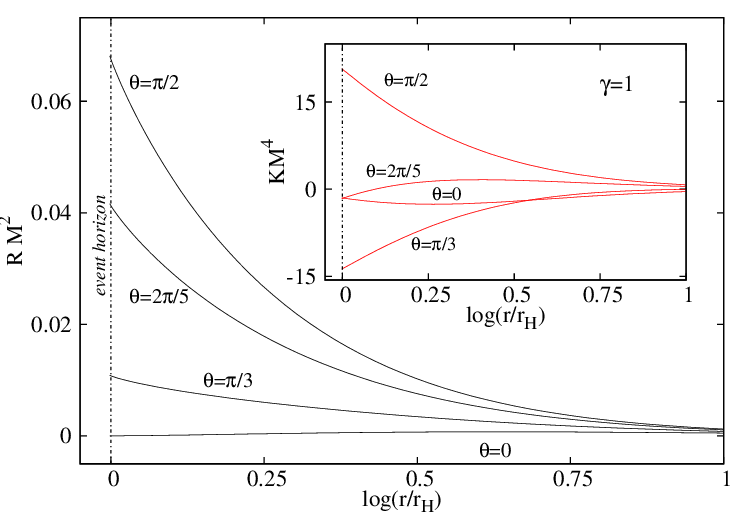}
		\end{subfigure}%
		\begin{subfigure}[b]{7.85cm}
			\centering 
\includegraphics[width=7.85cm]{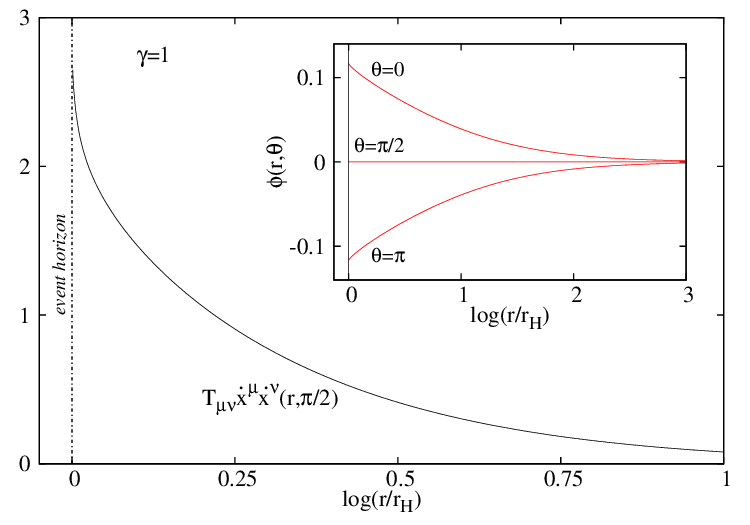}
		\end{subfigure}%
        }
%\\
%\\
\caption{\small  
{\it Left:}
The Ricci and Kretschmann scalars (in units of mass) are shown 
for a typical  $\gamma=1$ dyonic spinning BH
with $j=0.825$.
{\it Right:}
The energy density as measured by a unit energy particle
infalling along a radial geodesic  at $\theta=\pi/2$ 
%$T_{\mu \nu} \dot x^\mu \dot x^\nu$
is shown  for the same solution,
together with the scalar field profile 
at three different angles.
}
\label{pathology}
\end{figure}

%
These solutions are regular,
with finite Kretschmann  and Ricci scalars,
and also without a $pp$ curvature
singularity.
Furthermore, we have computed also the measure 
$T_{\mu \nu}\dot x^\mu \dot x^\nu$ 
(which gives the energy density measured by a unit energy particle
infalling along a radial geodesic),
with 
\begin{equation}
\dot t= \frac{1}{x^4}e^{-2 F_0} E H^2u, \quad
\dot x= e^{-F_1}\sqrt{\frac{e^{-2 F_0}E^2 u  }{x^2}
-\frac{x^2}{H^2}}, \quad
\dot \theta=0, \quad
\dot \varphi= \dot t W,
\end{equation}
 the tangent vector of a
timelike ingoing geodesic at $\theta=\pi/2$ parametrized by the proper time $\tau$.
As one can see in Figure \ref{pathology}, this quantity is also finite.
The figure exhibits the strong angular dependence of the scalar field, 
despite the fact that 
the dilaton charge vanishes, $D=0$. 

%%%%%%%%%%%%%%%%%%%%%%%%%%%%%%%%%%%%%%%%%%%%%%%%%%%
 %{\small \hspace*{3.cm}{\it  } }
%%%%%%%%%%%%%%%%%%%%%%%%%%%%%%%%%%%%%%%%%%%

We also  found that
all spinning extremal BHs  have an ergoregion, defined as the domain in which the norm of $\xi=\partial_t$ becomes positive outside the horizon. This manifests such solutions are not supersymmetric~\cite{Herdeiro:2000ap}.
The ergoregion is bounded by the event horizon and by the surface where
\begin{equation}
 g_{tt}=-e^{2F_0} \frac{x^4}{u H^2}+W^2e^{-2F_0}u \sin^2\theta =0 \ .
\end{equation}
As with  the extremal KN BH, this surface has a spherical topology 
and touches the horizon at the poles. 
Another noteworthy feature is that
these dyonic eBHs
do not possess the usual
north-south (reflection) symmetry,
a feature  also  encountered for the $\gamma=\sqrt{3}$
exact spininning, dyonic solution~\cite{Rasheed:1995zv,Larsen:1999pp}.

This family of
regular eBHs stops to exist for a critical configuration with 
$j\simeq 0.58722$,
which is denoted with $C$ in all plots in this work.
Although a continuation of this set 
 for smaller $j$ appears to exist, it
possesses all issues of the generic solutions
with $P^2 \neq Q^2$ and we shall not discuss it.

Related to the last observation, the 
construction of regular eBHs starting 
with the extremal 
EMd spherically symmetric  BHs with 
$\gamma=1$ was unsuccessful.
This limiting solution is known in closed form for general $P$ and $Q$~\cite{Kallosh:1992ii},
and extensively discussed in the literature, since it can be embedded in ${\cal N} = 4$
supergravity.
The $P=Q$ extremal limit of the solution in~\cite{Kallosh:1992ii} has a particularly simple form, with
\begin{eqnarray}
\label{RN}
ds^2=-\left(1-\frac{\sqrt{2} Q}{ r}\right)^2 dt^2+\frac{dr^2}{ (1-\frac{\sqrt{2} Q}{ r})^2 }+r^2
(d\theta^2+\sin^2 \theta d \varphi^2),~~
A=Q \cos\theta d\varphi-\frac{Q }{r} dt,~~ \phi(r)=0,
\end{eqnarray}
coinciding with the  Reissner-Nordstr\"om eBH.
Intriguingly, no  smooth
spinning generalizations of this configuration could be found,
even in the slowly rotating limit.

%%%%%%%%%%%%%%%%%%%%%%%%%%%%%%%%%%%%%%%%%%%%%
\section{The near horizon solutions}
\label{sec3}
%%%%%%%%%%%%%%%%%%%%%%%%%%%%%%%%%%%%%%%%%%%%%

%%%%%%%%%%%%%%%%%%%%%%%%%%%%%%%%%%%%%%%%%%%%%%%%%%%%%%
\subsection{The near horizon limit of the bulk solutions}
%%%%%%%%%%%%%%%%%%%%%%%%%%%%%%%%%%%%%%%%%%%%%%%%%%%%%%

A partial
understanding of (some of) the  solutions' properties,
in particular of the condition
(\ref{condition}),
can be
achieved by studying their near horizon 
limit.

The starting point here is to   % a new radial coordinate 
consider  the transformation 
$x =\lambda \sqrt{R}$,
$t=T/\lambda$
in (\ref{line_BHn})
and 
 take the limit
$\lambda \to 0$.
After several redefinitions, 
this results in a configuration of the generic form 
\cite{Astefanesei:2006dd}
\begin{eqnarray} 
\label{metric}
ds^2= v_1(\theta) 
\left(
      -R^2 dT^2+\frac{dR^2}{R^2}+\beta^2 d\theta^2
\right)
+v_2(\theta) \sin^2 \theta (d \tilde \varphi-K R dT)^2~,
\end{eqnarray} 
for the geometry, and 
\begin{eqnarray} 
\label{matter-nh}
A=A_\varphi (\theta) (d \tilde \varphi- K R dT)+q R dT,~~\phi \equiv \phi(\theta)~,
\end{eqnarray} 
for the matter fields.
The correspondence with 
the bulk expressions
(\ref{line_BHn}),
(\ref{expansion-eh})
is straightforward, with 
 \begin{equation}
 T=\frac{e^{\alpha}}{2r_H^2}t,~\tilde \varphi = \varphi -\Omega_H t,~
 v_1(\theta)= e^{2F_1(0,\theta)} {r_H^2},~
  v_2(\theta)= e^{-2(\alpha+F_1(0,\theta))} {r_H^2},~
 A_\varphi(\theta) =a_{\varphi{(0)}},~ 
  \phi(\theta)= \phi(0,\theta) ,~
\end{equation} 
and also 
$K=2 w_2e^{-\alpha} r_H^2$,
%$s= r_H^2 e^{-\alpha}$,
$q=a_{t(2)}$
(we recall $\alpha=(F_0-F_1)|_{x=0}$), while $\beta=1$
for the employed ansatz.

%%%%%%%%%%%%%%%%%%%%%%%%%%%%%%%%%%%%%%%
\subsection{Details on the problem} 
%%%%%%%%%%%%%%%%%%%%%%%%%%%%%%%%%%%%%%%%

%%%%%%%%%%%%%%%%%%%%%%%%%%%%%
\subsubsection{The equations} 
%%%%%%%%%%%%%%%%%%%%%%%%%%%%%%

Solving the
EMd equations for the Ansatz 
(\ref{metric}), (\ref{matter-nh})
is an interesting problem in itself, which  we shall consider  
in what follows,
ignoring for the moment the relationship with the bulk solutions.

The considered ansatz 
is written in term of four unknown functions
$\{ v_1(\theta),v_2(\theta);A_\varphi(\theta),\phi(\theta)\}$ 
and three constants $\{\beta, K,q\}$ .
However, there is a simple relation between the metric functions $v_1(\theta),v_2(\theta)$,
which is found as follows.
Let us define
\begin{eqnarray} 
v_2(\theta)=\frac{  {\cal S}^2(\theta)}{v_1(\theta)}.
\end{eqnarray} 
Then  the Einstein equation
$E_r^r+E_\theta^\theta=0$
implies that $ {\cal S}(\theta)$
solves the equation
\begin{eqnarray} 
\frac{1}{\sin^2\theta} (\sin^2\theta  {\cal S}')'
+(\beta^2-1) {\cal S}=0,
\end{eqnarray} 
where the prime denotes a derivative $w.r.t.$ 
the coordinate 
$\theta$,
with the solution
\begin{eqnarray} 
 {\cal S}(\theta)=s \frac{ \cos \beta(\theta-\theta_0)}{\sin \theta},
\end{eqnarray} 
where $\theta_0$ and $s$ are integration constants.
The physical solutions should possess no singularity
on the $z-$axis ($\theta=0,\pi$), $i.e.$
\begin{eqnarray} 
\label{regular}
 \lim_{\theta \to 0}\frac{g_{\varphi \varphi}}{g_{\theta \theta}}=\theta^2+\dots,~~
{\rm and}~~
\lim_{\theta \to \pi}\frac{g_{\varphi \varphi}}{g_{\theta \theta}}=(\pi-\theta)^2+\dots,~~
\end{eqnarray} 
which implies
%\begin{eqnarray} 
%\label{new-cond1}
$\theta_0=\frac{\pi}{2}$
and
$\beta=1$.
%~{i.e.}~~\bar S=S.
%\end{eqnarray} 
Therefore the following
 simple relation holds,
\begin{eqnarray} 
v_2(\theta)=\frac{s^2}{v_1(\theta)},%~~~{\rm and}~~\beta=1,
\end{eqnarray} 
with $s$ an %(undetermined) 
integration constant.
From (\ref{regular}), 
this constant
correspond to the value
taken by the function
$v_1$ at the poles of the $S^2$-sphere,
\begin{eqnarray} 
\label{new-cond}
s=v_1 (0)=v_1 (\pi).
\end{eqnarray} 
Therefore the EMd equations imply the following set of three
ODEs
\begin{eqnarray} 
\nonumber
&&
v_1''+v_1 
-\frac{3s^2 K^2 \sin^2 \theta}{4v_1}
-\frac{3v_1'^2}{4 v_1} 
+ v_1 \phi'^2
-e^{-2\gamma \phi}
\left(
(q-K A_\varphi)^2
+\frac{v_1^2 A_\varphi'^2}{s^2 \sin^2 \theta}
\right)=0,
\\
&&
\label{eqs}
A_\varphi''-\frac{s^2 K  \sin^2 \theta}{v_1^2}(q-K A_\varphi) 
-A_\varphi' \left(2 \gamma \phi'+\cot \theta - \frac{v_1'}{v_1} \right)=0,
\\
&&
\nonumber
\phi''+\cot \theta \phi'-\gamma\, e^{-2\gamma \phi} 
\left(
\frac{(q-K A_\varphi)^2}{v_1}-\frac{v_1A_\varphi'^2}{s^2 \sin^2 \theta}
\right) =0.
\end{eqnarray} 
In addition, there is an extra  (constraint)  Einstein equation
%which in  what follows  is not solved directly, %being a linear combination of 
%the eqs. (\ref{eqs}),
\begin{eqnarray} 
\label{constr}
-1+\frac{v_1'^2}{4v_1^2}
-\frac{\cot \theta v_1'}{v_1}
+\frac{s^2 K^2 \sin^2 \theta}{4v_1^2}
+\phi'^2
+\frac{e^{-2\gamma \phi} v_1}{s^2  \sin^2\theta}
  A_\varphi'^2
+\frac{e^{-2\gamma \phi}}{v_1  }
  (K A_\varphi-q)^2
 =0~.
\end{eqnarray} 
One can prove the a combination of the equations for $(A_\varphi,\phi)$
results in the
following first integral:
\begin{eqnarray} 
\label{FI}
\sin \theta ~\phi'+\frac{\gamma e^{-2 \gamma \phi}}{2s^2 K^2}\frac{v_1}{\sin \theta}
\frac{d}{d\theta}(K A_\varphi-q)^2=p_0~,
\end{eqnarray} 
with $p_0$ an integration constant.
As an additional observation, the integral of a suitable combination of Einstein equations
leads to the following interesting expression:
\begin{eqnarray} 
\label{reln}
\int_0^\pi d\theta
\bigg[
\frac{s^2 K^2 \sin^3 \theta}{2v_1^2}
+\frac{e^{-2\gamma \phi}\sin \theta}{ v_1}
\left(
  (K A_\varphi-q))^2 +\frac{v_1^2 }{s^2  \sin^2 \theta}A_\varphi'^2
\right)
\bigg]=2.
%~~{\rm with}~~U=W A_\varphi-q~.
\end{eqnarray}

For completeness, let us mention that the
equations of the model can be derived from
the following effective Lagrangian:
\begin{eqnarray} 
\label{Leff}
{\cal L}_{\rm eff}=
-\frac{\sin \theta v_1'^2}{2v_1^2}
+\frac{2\cos \theta v_1'}{v_1}
+\frac{s^2 K^2 \sin^3 \theta}{2v_1^2}
-2\sin \theta
\bigg[
\phi'^2
+e^{-2\gamma \phi}
v_1 \left(\frac{A_\varphi'^2}{s^2   \sin^2 \theta} -\frac{(K A_\varphi-q)^2}{v_1^2}\right)
\bigg]~,
\end{eqnarray}
which provides a bridge 
with the {\it rotating attractors} \cite{Astefanesei:2006dd}.
The advantage
of the latter approach is that one can directly
compute the physical
charges $J,Q$ and
(usually)
find the
entropy as a function of them.
In what follows we shall not pursue this direction; but we 
 have verified that 
it provides the same equations (\ref{eqs}),
since extremising the entropy
function is equivalent to the equations of motion while its extremal value corresponds to the
entropy $S=A_H/4$.
Moreover, 
 the same expressions 
(\ref{expr-Q}),
(\ref{expr-J})
are found
for $J$ and $Q$,
as expected.

 %%%%%%%%%%%%%%%%%%%%%%%%%%%%%%%%%% 
\subsubsection{Approximate expansion }
%%%%%%%%%%%%%%%%%%%%%%%%%%%%%%%% 

The small-$\theta$ expansion of the solutions reads
\begin{eqnarray} 
\nonumber
&&
v_1(\theta)= s+\frac{1}{2} 
\left(
-s+e^{-2\gamma \phi_{0(N)}}( 4u_{2(N)}^2+(q-P K)^2)
\right) \theta^2+\dots,~~
A_\varphi(\theta)=P
+u_{2(N)} \theta^2+\dots,
\\
\label{s0}
&&
\phi(\theta)=\phi_{0(N)}
+\frac{e^{-2\gamma \phi_{0(N)}} \gamma ( (q-P K)^2-4 u_{2(N)}^2 )}{4s}
\theta^2+\dots,~~
\end{eqnarray} 
in terms of three coefficients
%\begin{eqnarray} 
 $
 \{ P, u_{2(N)};\phi_{0(N)} \} \,.
 $
A similar expansion exists for
$\theta \to \pi$, 
with
\begin{eqnarray} 
&&
\label{spi}
v_1(\theta)= s+\frac{1}{2} 
\left(
-s+e^{-2\gamma \phi_{0(S)}}( 4u_{2(S)}^2+(q+P K)^2)
\right)
(\pi-\theta)^2+\dots,~~
\\
\nonumber
&&
A_\varphi(\theta)=-P+u_{2(S)} (\pi-\theta)^2+\dots,
~
\phi(\theta)=\phi_{0(S)}
+\frac{e^{-2\gamma \phi_{0(S)}} \gamma ( (q+PK)^2-4 u_{2(S)}^2 )}{4s}
(\pi-\theta)^2+\dots,~
\end{eqnarray} 
with two new free parameters  
%\begin{eqnarray} 
$
\{ u_{2(S)}; \phi_{0(S)} \}.
$
%\end{eqnarray} 

Additionally, the first integral 
(\ref{FI})
results in
\begin{eqnarray} 
e^{-2\gamma  \phi_{0(N)}}u_{2(N)}(P K-q))
=
e^{-2\gamma \phi_{0(S)}}u_{2(S)}
(q+P K)),
\end{eqnarray} 
which can be used to  express 
%$e.g.$
$u_{2(S)}$
in terms of other parameters.

%%%%%%%%%%%%%%%%%%%%%%%%%%%%%%%%% 
\subsubsection{Properties of the bulk
eBHs from the near horizon solutions}
%%%%%%%%%%%%%%%%%%%%%%%%%%%%%%% 

The knowledge of the near-horizon solution
allows to compute the electric and magnetic charges
as well as 
the angular momentum 
and the horizon area of the corresponding bulk eBHs.
None of the other quantities which enter
the thermodynamics ($e.g.$ ADM mass)
can be computed at this level.

The magnetic charge has the simple 
expression
\begin{eqnarray} 
\label{chargeP}
P=\frac{1}{2} \left (
A_\varphi(0)-A_\varphi(\pi)
\right).
 \end{eqnarray} 
 The situation with the electric
 charge and angular momentum is
 more complicated,
 and we compute that by using the 
 the relations
(\ref{Q}),
(\ref{J}).
This results in 
the following expressions 
\begin{eqnarray}
&&
\label{chargeQe}
Q= \int_0^\pi d \theta~
\frac{ S  \sin \theta}{2v_1} 
  e^{-2\gamma \phi}
(q-K A_\varphi)~,
\\
&&
\nonumber
\label{chargeJ}
J=\int_0^\pi d \theta~
\frac{s}{2K}
\left(
\frac{s^2 K^2 \sin^3 \theta}{4v_1^2}
+\frac{\sin \theta e^{-2\gamma \phi}}{v_1}K A_\varphi (K A_\varphi-q)
\right).
\end{eqnarray}
By using the  equations (\ref{eqs}),
one can show that
the integrand for both
$Q,J$ can be
written as a total derivative, 
\begin{eqnarray}
\label{expr-Q}
Q=\int_0^{\pi  } d\theta
\left(
\frac{v_1}{2s} 
\frac{e^{-2\gamma \phi}}{K \sin \theta }A_\varphi'
\right)'=
\frac{e^{-2\gamma \phi}}{2 K \sin \theta }A_\varphi' \bigg |_0^\pi =-
\frac{1}{K}
\left(
e^{-2\gamma \phi (0)} u_{2(0) }
+e^{-2\gamma \phi (\pi)} u_{2 (S)}
\right),
\end{eqnarray}
and
\begin{eqnarray}
\nonumber
J &=& \int_0^{\pi  } d\theta
\left[
\frac{s}{8K}(\frac{\sin \theta v_1'}{v_1}-2\cos \theta))
-\frac{q}{K}
\left(
\frac{v_1}{2s} 
\frac{e^{-2\gamma \phi}}{K \sin \theta }A_\varphi'
\right)
-\frac{1}{4K^2}
\frac{v_1}{s}\frac{A_\varphi'}{\sin \theta}
e^{-2\gamma \phi}(K A_\varphi-q)
\right]'
\\
\label{expr-J}
&=&\frac{ 1}{2K}( s -2q Q )~,
\end{eqnarray}
 where  in the last step we have used the first integral
  (\ref{FI}).
The event horizon area is fixed by the
input constant $s$:
\begin{eqnarray} 
A_H=4\pi s.
\end{eqnarray} 
Thus, 
$J$, $P$ and $A_H$
are fixed by data at the poles
of the horizon. The 
horizon angular momentum reads
\begin{eqnarray} 
\label{JH_jose}
J_H=
\frac{s^3 K }{8}\int_0^\pi d\theta
\frac{\sin^3 \theta}{v_1^2}.
\end{eqnarray}

%%%%%%%%%%%%%%%%%%%%%%%%%%%%%
\subsubsection{Symmetries} 
%%%%%%%%%%%%%%%%%%%%%%%%%%%%%%

The considered system  
has a number of symmetries: transformations leaving  Eqs. (\ref{eqs}) invariant.
In what follows we divide them 
into two classes, depending on the presence of the scalar field or not.

Let us start with the
{\it general symmetries}: 
\begin{enumerate}[label=(\roman*)]
\item
Invariance under changing the
 sign for both $ A_\varphi$ and $q$,
\begin{eqnarray} 
\label{symm1}
{\bf g_1:}~~~
 A_\varphi \to -  A_\varphi,~~q \to -q \implies (P\to -P,~Q  \to -Q)~.
\end{eqnarray} 

\item
Invariance under changing the
 sign for both $ A_\varphi$ and $K$,
\begin{eqnarray} 
\label{symm1n}
 {\bf g_2:}~~~A_\varphi \to -  A_\varphi,~~K\to -K\implies (P\to -P,~J  \to -J)~-
\end{eqnarray} 

\item
Eqs. 
(\ref{eqs})
possess 
the scaling symmetry
\begin{eqnarray} 
\label{scaling}
 {\bf g_3:}~~~A_\varphi \to \lambda  A_\varphi,~~s \to \lambda s,~~{\rm and}~~
K \to \frac{K}{\lambda},
\end{eqnarray}  
with $\lambda$ an arbitrary positive parameter. This symmetry, however, 
is $not$ present for 
regular solutions, since would imply
the occurrence of conical singularities
(note that $v_1$ does not change under (\ref{scaling}), 
and thus
$v_1(0)\neq s$).

 \item 
Another scaling symmetry  is
\begin{eqnarray} 
\label{symm2}
{\bf g_4:}~~~
A_\varphi\to A_\varphi+\lambda,~~q \to q+K \lambda.
\end{eqnarray}  
which is also not present for physical solutions, since it violates
the condition
$A_\varphi(0)+A_\varphi(\pi)=0.$

\end{enumerate}

The {\it specific symmetries} are:
\begin{enumerate}[label=(\roman*)]
\item
The system remain invariant under a scaling of
the line-element
together with a shifting of the scalar field,
with 
 \begin{eqnarray} 
\label{scaleS}
{\bf s_1:}~~~
(v_1,S) \to e^{-2\gamma \phi_0} (v_1,S),~ 
\phi \to \phi+\phi_0
\implies
(A_H,J,Q ) \to e^{-2\gamma \phi_0}
(A_H,J,Q )~.
\end{eqnarray} 
where $\phi_0$ is an arbitrary number.

\item 
Finally, there is yet another
scaling symmetry of the matter system, which does not affect the geometry
 \begin{eqnarray} 
\label{symm3}
{\bf s_2:}~~~(A_\varphi,q) \to e^{ \gamma \phi_0} (A_\varphi,q)  ,~~
\phi \to \phi+\phi_0,
\implies 
Q \to e^{ -\gamma \phi_0} Q ,~
P\to e^{\gamma \phi_0} P.
\end{eqnarray} 
The physical relevant quantities 
are those   which
remain invariant under the
above symmetries,
like
$PQ/J$
and $J/A_H$.

\end{enumerate}

%%%%%%%%%%%%%%%%%%%%%%%%%%%%%%%%%%%%%%%%%%%%%%%%%  
 %\appendix
\subsection{Perturbative results}
%\setcounter{equation}{0}
%\renewcommand{\theequation}{A.\arabic{equation}}
%%%%%%%%%%%%%%%%%%%%%%%%%%%%%%%%%%%%%%%%%%%%%%%%%%%%%%% 

\subsubsection{
Charging the NHEK geometry.
The $P=Q$ condition.
}
 
Since the near horizon extremal Kerr (NHEK) vacuum configuration is a solution
of the EMd equations for any $\gamma$,
one can attempt to construct a perturbative solution around it.
 
In solving in closed form the eqs. (\ref{eqs}), 
it is convenient to use a 
new coordinate
\begin{eqnarray} 
\label{u}
u=\cos \theta.
\end{eqnarray} 
Then by using a perturbative ansatz with  
\begin{eqnarray} 
&&
v_1(u)= \sum_{ k\geq 0} \epsilon^k v_{1k}(u),~~
A_\varphi (u)= \sum_{ k\geq 0} \epsilon^k 
A_{\varphi (k)}(u),~~
\phi (u)= \sum_{ k\geq 0} \epsilon^k \phi_k(u),~~
 \\
 \nonumber
 &&
{\rm and}~~~
K= \sum_{ k\geq 0} \epsilon^k  K_k ,~~
 q= \sum_{ k\geq 0} \epsilon^k q_k ,~~
\end{eqnarray}  
(with $\epsilon$ an infinitesimally small parameter),
the equations
 can easily be solved order by order.
The zeroth order in the expansion above corresponds to 
the NHEK solution, with:
\begin{eqnarray} 
v_{10}(u)=a^2(1+u^2),~~A_{\varphi (0)}(u)=0,~~
 \phi_0(u)=0,~~
s=2a^2 ,~~K_0=1,
\end{eqnarray} 
with $a>0$ an arbitrary parameter fixing the length scale of the problem.
A number of integration constants in the expression of
 the higher order terms
are fixed by imposing that the solution is smooth everywhere, in particular without conical singularities.
For the scalar field we impose
$\phi=0$
for $A_\varphi=0$.
Then, 
at  each order $k>0$ in perturbation theory,
there is a single free constant of integration, $P_k$
which fixes the $k$-th order contribution to the total
magnetic charge.
The Kaluza-Klein case 
 turns out  to be special, 
the factor $\gamma^2-3$
entering in the 
perturbation equations in various places, which results in alternative choices of the integration constants.

In the general-$\gamma$ case,
the solution to fourth order in perturbation theory\footnote{No obstacle appears to exist in extending
(\ref{solq1})-(\ref{solq3})
to arbitrary order. In fact, 
we have solved the 
equations up to eight order, however
without being possible to identify a general pattern. } reads
\begin{eqnarray} 
&&
\nonumber
v_{11}(u)=0,~~
v_{12}(u)=  P_1^2 (1-u^2),~~
v_{13}(u)= 2 P_1 P_2 (1-u^2),~~~~
\\
\label{solq1}
&&
v_{14}(u)=   (1-u^2)
\left(
P_2^2+2 P_1 P_3
+
\frac{\gamma^2 P_1^4}{6  a^2 }
\frac{ (1-u^2) }{ (1+u^2) }
\right),~~~~
\end{eqnarray}  
\begin{eqnarray} 
&&
\label{solq2}
A_{\varphi (1)}(u)=  P_1 \frac{1+2u-u^2}{ 1+u^2 },~~
A_{\varphi (2)}(u)=  P_2 \frac{1+2u-u^2}{ 1+u^2 },~~
\\
&&
\nonumber
A_{\varphi (3)}(u)=    \frac{u^2+ 2u-1}{ 1+u^2 } 
\left( 
P_3-\frac{P_1^3}{a^2}\frac{1-u^2}{1+u^2}
\right)
+\frac{\gamma^2 P_1^3(1-u^2)(u^2-2u-1)}{3a^2 (1+u^2)^2},
\\
&&
\nonumber
A_{\varphi (4)}(u)=  P_4 \frac{1+2u-u^2}{ 1+u^2 )}
+\frac{3P_1^2 P_2(1-u^2)}{ a^2(1+u^2)^2}
\left(
 3-2u+u^2
 -\frac{\gamma^2 }{3}
( 3+2u+u^2) 
\right),
\end{eqnarray} 
\begin{eqnarray} 
&&
\label{solq3}
\phi_{1}(u)= 0,~~
\phi_{2}(u)= -\frac{\gamma P_1^2 u}{ a^2 (1+a^2)},~~
\phi_{3}(u)= -\frac{2\gamma P_1 P_2 u}{ 2a^2 (1+u^2)},~~
\\
&&
\nonumber
\phi_{4}(u)= \frac{\gamma u}{ a^4 (1+u^2)^2}
 \left(
P_1^4 (u^2-1)+a^2 (1+u^2)(P_2^2+2 P_1 P_3)
 \right)
+\frac{\gamma(3-\gamma^2)P_1^4}{3a^4}\arctan u,~~
\end{eqnarray}  
and
\begin{eqnarray} 
&&
 K_1= K_2=K_3=0,~
 K_4=\frac{(2\gamma^2-3)P_1^4}{ 6a^4},
 \\
 &&
 q_1= q_2= 0,~q_3=\frac{(3-2\gamma^2)P_1^3}{ 3a^2},~
q_4=\frac{(3-2\gamma^2)P_1^2 P_2}{  a^2}.
\end{eqnarray} 

The following relations follow\footnote{A similar solution exist with  $P=-Q$.} 
\begin{eqnarray}  
&&
P=\frac{1}{2}\sum_{k\geq 1} P_k q^k= Q,~~
%\\
%&&
{\rm and}~~
J=a^2+\frac{(2\gamma^2-3)P_1^4 }{6 a^2}\epsilon^4+\dots~,
\end{eqnarray} 
with $P,~Q$, $J $
computed from (\ref{chargeP}), (\ref{chargeQe}) and  (\ref{chargeJ}), 
respectively. In particular, observe how smoothness enforces the $P=Q$ condition.
Also, it is interesting to notice that, up to this order\footnote{
We have found that the next order contribution to area in
(\ref{area-pert}) is proportional to $P^3 Q^3/J^2$.},
the horizon area of the solutions can be written as
\begin{eqnarray}  
\label{area-pert}
&&
A_H= 8\pi J+\frac{4 \pi (3-2\gamma^2)}{3}\frac{P^2 Q^2}{J}.
\end{eqnarray}

%%%%%%%%%%%%%%%%%%%%%%%%%%%%%%%%%%%%%%%%%%%%%%%%%%
\subsubsection{A small-$\gamma$  result
}
%%%%%%%%%%%%%%%%%%%%%%%%%%%%%%%%%%%%%%%%%%%%%%%%%% 
Further evidence for the condition $P=Q$
comes from a perturbative study in $\gamma$ 
around  
%The solution for $\gamma=0$ 
%corresponds to 
the near horizon limit
of the extremal dyonic KN BH.
This $\gamma=0$ solution 
reads
%\cite{Bardeen:1999px}
%
 \begin{eqnarray} 
\label{KN1n}
v_1(u)= r_0^2\left(1-\frac{a^2}{r_0^2} (1-u^2) \right ),~~
A_\varphi(u)= \frac{1}{1-\frac{a^2}{r_0^2} (1-u^2) } 
\left(
Pu -\frac{ q Q  M}{r_0^2}(1-u^2) 
  \right)~,~~\phi=0~,
\end{eqnarray} 
in terms of the parameters $Q,P $ and $a$.
Also,
\begin{eqnarray} 
\label{KN2}
q = \frac{ Q (Q^2+P^2) }{ r_0^2},~~ 
s=r_0^2,~~K=\frac{2aM}{r_0^2}~,
~~{\rm with}~~
 r_0=\sqrt{2a^2+Q^2+P^2 },~~
    M=\sqrt{ a^2+Q^2+P^2 },
\end{eqnarray} 
where $a=J/ M$. 

One can  try to construct a perturbative solution in $\gamma$
around the solution (\ref{KN2}).
To lowest order, one takes 
\begin{eqnarray} 
\phi (u)=\gamma \phi_1 (u),
\end{eqnarray} 
a simple computation resulting in 
\begin{eqnarray} 
&&
 \phi_1 (u)=p_0+p_1 \arctanh(u)
 -\frac{1}{(2a^2+Q^2+P^2)^2}
 \bigg[
 4q PQ \sqrt{a^2+Q^2+P^2}\arctanh(u)
 \\
 \nonumber
 &&
 -2 PQ (P^2+Q^2)\arctan (\frac{a u}{\sqrt{ a^2+Q^2+P^2 }})
 +\frac{1}{2}(Q^4-P^4)\log(\frac{1-u^2}{Q^2+P^2 +a^2(1-u^2)})
 \\
  \nonumber
 &&
 +
 \frac{(2a^2+Q^2+P^2)((Q^2+P^2)( a^2+Q^2+P^2)-2 aPQ \sqrt{ a^2+Q^2+P^2 }u)}{( Q^2+P^2+a^2 (1+u^2))}
 \bigg],
\end{eqnarray} 
with $p_0$, $p_1$
integration constant.
The condition for a regular
scalar field  at $u=1$
implies
\begin{eqnarray}
p_1=\frac{P^4-Q^4+ 4a PQ \sqrt{a^2+Q^2+P^2}}{(2a^2+Q^2+P^2)^2}
\end{eqnarray} 
The scalar field is regular also at 
$u=-1$ 
if the magnetic and electric charges are equal, 
\begin{eqnarray}
P^2=Q^2~,
\end{eqnarray}
only.

Choosing $P=Q$ and $p_0=0$,
the simplified expression\footnote{The choice $P=-Q$ results
in $ \phi_1\to - \phi_1$.} of the 
scalar field reads
\begin{eqnarray}  
 && 
 \phi_1 (u)=
  \frac{Q^4}{(a^2+Q^2)^2}
  \arctan(
  \frac{a u}{\sqrt{a^2+2Q^2}}
  )
  +\frac{aQ^2 \sqrt{a^2+2Q^2}u}{(a^2+Q^2)( 2Q^2+a^2 (1+u^2)}~,
 \end{eqnarray}
 which is regular everywhere.

%%%%%%%%%%%%%%%%%%%%%%%%%%%%%%%%%%%%%%%%%%%%%%%%%%
\subsubsection{
The case of slowly rotating  dyons }
%%%%%%%%%%%%%%%%%%%%%%%%%%%%%%%%%%%%%%%%%%%%%%%%% 
In principle, the rotating configurations
could be found by considering a perturbative
approach 
 in terms of the parameter  $K$
 in (\ref{metric})
around the static EMd dyonic near horizon solutions.
Employing again the $u$-coordinate (\ref{u}),
  one writes
\begin{eqnarray} 
&&
\nonumber
\phi (u)=\phi_0+K \phi_1 (u)+\dots,~~
v_1 (u)= 2e^{-2 \gamma \phi_0}P^2+K v_{11} (u)+\dots,~~
\\
\label{pert}
&&
A_\varphi (u)=P u+  K A_{\varphi,1} (u)+\dots,~~
q=P+K q_1+\dots, 
\end{eqnarray} 
the $K=0$ case   corresponding to  
the static solution.
Then one finds from the Maxwell equations
 \begin{eqnarray} 
v_{11}(u)= 2e^{-2 \gamma \phi_0}P (Pu+2\gamma P \phi_1(u)
 -A_\varphi' (u) )+v_{110},
 \end{eqnarray}
 with $v_{110}$ a constant of integration.
In the next step, the  equation for the scalar field  is
integrated for 
$\phi_1 (u)$.
Its  general-$\gamma$ expression is exceedingly complicated;
however, the case $\gamma=1$
is relatively simple, with
 \begin{eqnarray} 
\phi_1 (u)=
c_0 u \arctanh u
 +\frac{1}{3}u \log (1-u^2)+ u c_1-c_2
,~~{\rm where }~~c_0=\frac{P(q_1+2 P c_2)-\frac{1}{2}e^{2 \phi_0}v_{110}}{2P^2}~.
\end{eqnarray}
One can verify that,
for any choice of the integration constants
$c_1,~ c_2,\phi_0,~v_{110}$,
the scalar 
$\phi_1 (u)$
is divergent at $u = 1$ and/or
$u = -1$. 
This result partially explains the numerical difficulties
encountered when attempting to construct
bulk BHs starting with an static extremal solution  as initial guess.

%%%%%%%%%%%%%%%%%%%%%%%%%%%%%%
\subsection{Nonperturbative numerical results}
%%%%%%%%%%%%%%%%%%%%%%%%%%%%%%% 
Equations  (\ref{eqs})
have been solved numerically as a  
multishooting problem,
by using both a standard F90 solver
and a custom  code implemented in Julia, 
 with agreement to high accuracy.
 In this approach, one starts with the expansion  (\ref{s0}) 
 close to $\theta=0$
 and integrates towards  $\theta=\pi$,
 adjusting for the parameters in (\ref{s0}), together with
 $\{ s, q, P,K  \}$,
 such that the expansion  
 (\ref{spi})
 ĩs satisfied.
 A number of solutions were also derived by using a different software, which involves a Newton–Raphson method \cite{Ascher:1979iha,Ascher:1981bph}.
 
\begin{figure}[h!]
\hbox to\linewidth{\hss%
	\resizebox{9cm}{7cm}{\includegraphics{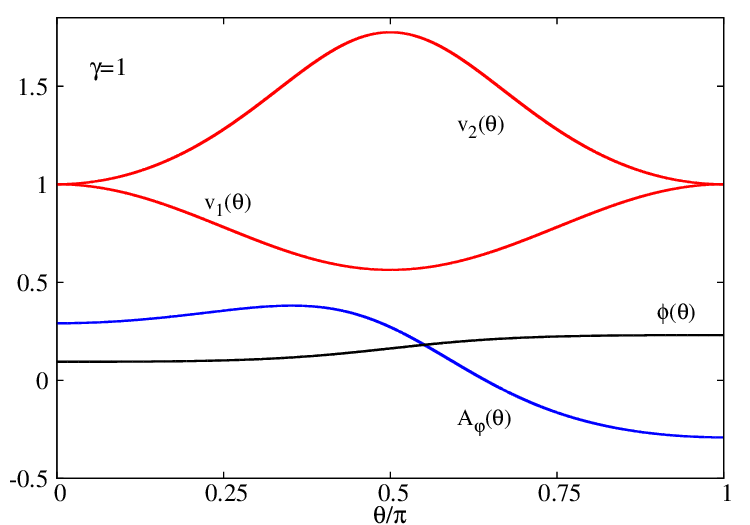}} 
 \resizebox{9cm} {7cm}{\includegraphics{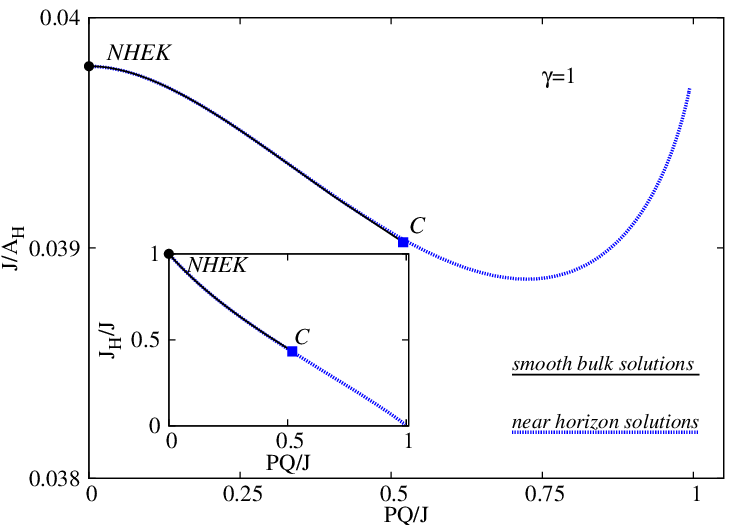}}  
\hss}
%\\
%\\
\caption{\small  
{\it Left:}
The profile of a typical rotating
near horizon solution with $\gamma=1$.
{\it Right:} A comparison between the results for 
extremal  (smooth)  BH solutions (red curve)
and near horizon configurations (blue dotted curve).
The two curves coincide up to the critical configuration $C$, only.
}
\label{profile}
\end{figure}

 The profile of a typical solution is shown in Figure \ref{profile}, left panel
 (note there the absence of 
a reflection symmetry $w.r.t.$
an equatorial plane).

%%%%%%%%%%%%%%%%%%%%%%%%%%%%%%%%%%%%%%%%%%%%%%%%%%%
 {\small \hspace*{3.cm}{\it  } }
%%%%%%%%%%%%%%%%%%%%%%%%%%%%%%%%%%%%%%%%%%%
% 

A number of relevant quantities invariant under the scalings (\ref{symm3})
are shown in Figure \ref{profile} (right panel).
There, for comparison, we include also the corresponding data for the bulk (smooth) solutions. 
As one can see, although the two curves coincide for small enough value of $PQ/J$,
this holds only up to the critical configuration $C$,
where smooth bulk solutions cease to exist.

One remarks that
not all near horizon solutions are also
realized as global (asymptotically flat)  eBHs\footnote{This should not be a surprise,
 a similar behaviour being  found in other models \cite{Chen:2008hk,Blazquez-Salcedo:2013muz}.
}.
That is, the set of near horizon solutions emerges from the NHEK configuration (which has 
$PQ/J=0$, $J_H=J$ and $J/A_H=1/(8\pi)$),
coinciding with the data from smooth bulk eBHs up to the point $C$,
%after deviating from the bulk eBHs at the point $C$,
and ending in a  configuration with $J_H\simeq 0$
and $PQ\simeq J$.
We conjecture the limiting configuration to be singular, 
the Kretschmann scalar diverging at the poles of the sphere.

%%%%%%%%%%%%%%%%%%%%%%%%%%%%%%%%%%%%%%%%%%% 
\section{ Conclusions}
\label{sec4}
%%%%%%%%%%%%%%%%%%%%%%%%%%%%%%%%%%%%%%%%%%%

We have considered charged rotating dyonic BHs in four dimensional Einstein-Maxwell-dilaton theory. 
Restricting to extremal configurations
and a stringy value of the dilaton coupling
constant $\gamma=1$, we
have found that the only solutions which are regular on and outside the horizon
are those with equal (absolute) values for the electric and magnetic charges.
A partial understanding of this feature
is achieved when studying the decoupled near-horizon limit of the bulk solutions.

Our approach has been numerical. 
In particular, we have introduced a suitable metric parametrization and a numerical scheme,   
well suited for 
the numerical construction of eBHs. 
As for near horizon solutions,   we have used  both analytical and numerical
methods, a closed form perturbative  solution being reported. 
For both eBHs and near horizon solutions,
the approach in this work can easily be adapted to other models.

In supersymmetric theories, eBHs are often supersymmetric.
 The spinning solutions 
of the  model (\ref{action}) with $\gamma=1$ cannot be consistently embedded in $D=4$, $\mathcal{N}=4$ supergravity, since rotation turns on an additional axionic term {(see \cite{Chow:2014cca} for rotating black holes in 4D supergravity and \cite{dosSantosCostaFilho:2025ibq}  for rotating extremal black holes/attractors with axionic hair)}. 
 At the same time, for the known generalizations
 of the extremal KN BH, the magnetic and electric charges are independent parameters.
 It would be interesting to see if the condition (\ref{condition})
can also be circumvented
for extensions of the model  (\ref{action})
with other matter fields ($e.g.$ an axion \cite{Shapere:1991ta,Cremonini:2024eog,Gibbons:1982ih})
or in the presence of a dilaton potential \cite{Anabalon:2013qua,Astefanesei:2019qsg}.

%%%%%%%%%%%%%%%%%%%%%%%%%%%%%%%%%%%%%%%%%%%%%%%%%%%%%%%%%%%%%%%%%%%%%%%%%%%%%%%%%%%%%%%%%%%%%%%%%%%%%%
\section*{Acknowledgements}
%%%%%%%%%%%%%%%%%%%%%%%%%%%%%%%%%%%%%%%%%%%%%%%%%%%%%%%%%%%%%%%%%%%%%%%%%%%%%%%%%%%%%%%%%%%%%%%%%%%%%%

We thank D. Astefanesei for useful discussions.
This work is supported by CIDMA under the Portuguese Foundation for Science and Technology (FCT, https://ror.org/00snfqn58) Multi-Annual Financing Program for R\&D Units, grants UID/4106/2025 and UID/PRR/ 4106/2025, as well as the projects: Horizon Europe staff exchange (SE) programme HORIZON-MSCA2021-SE-01 Grant No. NewFunFiCO-101086251;  2022.04560.PTDC (\url{https://doi.org/10.54499/2022.04560.PTDC}) and 2024.05617.CERN (\url{https://doi.org/10.54499/2024.05617.CERN}). E.S.C.F. is supported by the FCT grant PRT/BD/  153349/2021 (\url{https://doi.org/10.54499/PRT/BD/153349/2021}) under
the IDPASC Doctoral Program.
JLBS gratefully acknowledges support from MICINN projects 
PID2021-125617NB-I00 ``QuasiMode'' and CNS2023-144089 ``Quasinormal modes''. The work
of K.U. is supported by the Grant “Fujyukai” from Iwanami Shoten, Publishers.

%!!!!!!!!!!!!!!!!!!!!!!!!!!!!!!!!!!!!!!!!!!!!!!!1
%!!!!!!!!!!!!!!!!!!!!!!!!!!!!!!!!!!!!!!!!!!!!!!!1
 \bibliographystyle{unsrt}

\bibliography{biblio}

\end{document}